\documentclass[mathleft
]{an}
\usepackage{graphicx}
\usepackage{times}
\begin{document}

\Pagespan{789}{}
\Yearpublication{2006}%
\Yearsubmission{2005}%
\Month{11}%
\Volume{999}%
\Issue{88}%

\title{A composite K-band Luminosity Function for Cluster Galaxies}

\author{R. De Propris\inst{1}\fnmsep\thanks{Corresponding author:
  \email{rdepropris@ctio.noao.edu}\newline}
\and  D. Christlein\inst{2}
}
\titlerunning{Cluster K-band LF}
\authorrunning{R. De Propris \& D.Christlein}
\institute{
Cerro Tololo Inter-American Observatory, La Serena, Chile
\and 
Max Planck Institute for Astrophysics, Garching, Germany
}

\received{}
\accepted{}
\publonline{}


\abstract{%
  We present a composite K-band luminosity function for 10 clusters
  at low redshift, where member galaxies are identified from an 
  existing spectroscopic survey (the 2dF galaxy redshift survey).
  Our kinematically selected K-band luminosity function is well 
  fitted by a Schechter function with $M^*_K=-24.50\ +\ 5\log h$
  and $\alpha=-0.98$ over $-27 < <M_K\ -5\log h < -22$. This is
  very similar to the 2dF field value and suggests that the integrated
  mass accretion history of galaxies does not vary strongly with
  environment}

\maketitle

\section{Introduction}

Clusters of galaxies are important for studies of galaxy formation
and evolution because they contain a {\it volume-limited} sample of
galaxies observed {\it at the same cosmic epoch} and therefore provide
a snapshot of galaxy properties in a homogeneous (high-density) 
environment to lookback times approaching 2/3 of the Hubble time.
Clusters are, however, special environments, which are known to affect
the morphology and star formation history of their members (e.g., Dressler
1980). If we wish to use cluster galaxies as evolutionary probes we need
to understand the effects of the cluster environment, in order to relate
our findings to the more general evolution of field galaxies. This has
recently become possible with large redshift surveys (such as 2dF and
the SDSS) which allow us to consider large samples of clusters spanning
a wide range of properties, and to extend the analysis to the lower
density groups and the general field.

The main conclusions of these studies are: (i) the optical luminosity
function of galaxies is independent of the cluster environment; there is
little variation between the field and clusters; (ii) star-forming galaxies
have the same luminosity function in all environments, while quiescent
objects show a significant difference, in that clusters contain a population
of dwarf ellipticals which is missing in the general field. The most
likely explanation for this involves the systematic suppression of 
star formation in low luminosity spirals and irregulars as they infall
into clusters and their transformation into dwarf ellipticals on very
short timescales, while giant galaxies appear to have resided in the
cluster environment for very large epochs (Lewis et al. 2002, Christlein 
\& Zabludoff 2003; De Propris et al. 2003, 2004; Gomez et al. 2003;
Balogh et al. 2004; Popesso et al. 2006).

This of course leaves some questions open: what is the role of mergers
in the environmental transformation of galaxies ? How much does the mass
of galaxies change after star formation is suppressed ? Are there any 
detectable differences between field and clusters, owing to their very
different densities and (presumably) merger histories ? What is the relative
role of star formation and mergers in assembling galaxies ? 

In order to answer these questions we need to study mass-selected samples of 
galaxies, to probe the stellar mass assembly history directly, and compare to
observations in bluer bands to understand the cumulative effect of star formation. 
Observations in the $K-$ band, which provide a measure of stellar masses (Gavazzi 
et al. 1996; Bell \& de Jong 2001), while minimally affected by star formation 
or dust, allow us to achieve these aims.

The purpose of this paper is to present preliminary conclusions from a
subset ($\sim 15\%$) of the observations of an ongoing infrared imaging
survey. We will assume the WMAP cosmology with $\Omega_M=0.73$ and 
$\Omega_{\Lambda}=0.27$ and H$_0=100$ km s$^{-1}$ Mpc$^{-1}$.

\section{Observations}

To this end, we began an observing campaign to image all clusters originally
studied by De Propris et al. (2003) in the $K$-band. These clusters have 
highly complete membership information, to the limits of the 2dF survey,
at least three (and sometimes more) photometric bands and the spectra can be
used to derive crude information on the galaxies' star formation histories.

Since the mean redshift of the clusters is $<z>=0.07$ these objects span large
areas on the sky. We therefore need to obtain $K$-band data over regions 
approaching one degree on the sky or more. The 2MASS data are not serviceable
for our purposes, as the imaging is too shallow, the star-galaxy separation
erratic and inconsistent, and the photometry inadequate for galaxies other than
local ones, owing to the very short exposures on a 1.3m telescope. We decided
to carry out deep imaging on 4m-class telescopes, with an integration time of
300s, to achieve sufficient depth, especially for the low surface brightness
regions of galaxies, to obtain a fair estimate of the total luminosity of
galaxies.

Observations for the first ten clusters were carried out with the ISPI array
on the CTIO 4m telescope in 2007 and 2008 (a total of ten nights, although 7
were lost to weather). The ISPI array covers $\sim 10.5'$ on the sky and we
obtained mosaics of ISPI images to image the extent of our clusters out to
their Abell radius (typically about one degree on the sky). 

The data consist of three coadded 20s exposures at each pointing, dithered in 
a five-point pattern with $20''$ steps, to obtain a total integration time of
300s in the inner $10'$ of each ISPI field. The data were reduced in the usual
way (for infrared data): after flatfielding and dark subtraction, we removed
a running median of imaging frames, to deal with the sky subtraction, and
the carried out alignment and corrections for the field distortions, 
with final co-addition of all images to produce a single frame.

Photometry was carried out using the standard Sextractor routines (Bertin \&
Arnouts 1996), and we calibrated our data directly on 2MASS stars (Jarrett et
al. 2000) present in our images. We used the Sextractor `stellarity' parameter 
and matched to the higher resolution data in the Supercosmos database (Hambly
et al. 2001) to separate stars and galaxies. We finally matched all galaxies
to the redshift database of the 2dF survey (Colless et al. 2001), as well as
all other available redshift data in the NED database (principally data from
the SDSS and the MGC surveys) for our observed fields.

We used the same approach as in De Propris et al. (2003) to correct for 
spectroscopic incompleteness in our K-band photometric survey, and its
match to existing spectroscopic surveys. For each absolute magnitude bin
we computed the fraction of members, non members, and galaxies whose
redshift is unknown, and applied a statistical correction to derive the
likely number of cluster members in each bin. For galaxies brighter than
$K=14$ the spectroscopic data are nearly 100\% complete, while the
spectroscopic completeness drops rapidly fainter than $K=15$. We plan 
to eventually exploit the deep photometry and partial redshift information
to study the dwarf galaxy population in greater detail.

We derived the composite luminosity function for our 10 clusters following
the method outlined in Colless (1989) and fitted this with a Schechter
function using a maximum-likelihood technique. Figure 1 shows the derived
LF and the error ellipses. The best fit values are $M^*_K=-24.50\ +5\log h$
and $\alpha=-0.98$. These are in good agreement with the values derived 
for Coma by De Propris et al. (1998) and by Lin et al. (2004). They are
also in very good agreement with the {\it field} value from the 2dF/2MASS
data of Cole et al. (2001). 

\begin{figure}
\includegraphics[width=3in,height=3in]{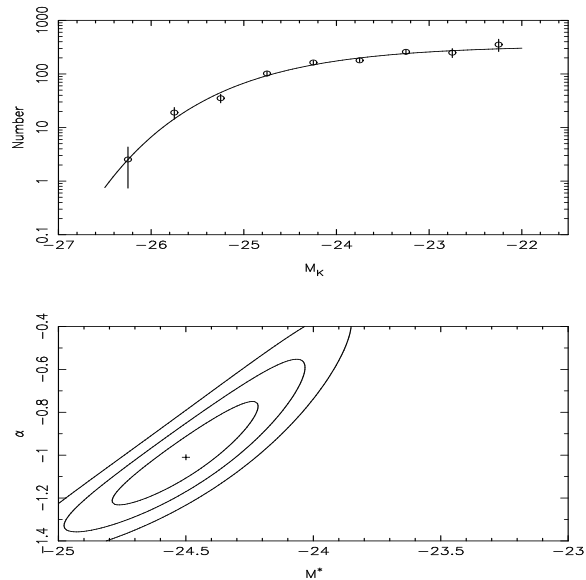}
\caption{The composite $K$-band LF for 10 clusters, best fit (top panel) and
error ellipse (bottom panel).}
\label{label1}
\end{figure}

\section{Discussion and Future Work}

The $K$-band LF we derive for cluster galaxies differs from other work (except
De Propris et al. 1998) in that we identify cluster members spectroscopically
and we therefore know the membership status of nearly every galaxy in our images.

Our LF is comparable in precision to the much larger study of Lin et al. (2004)
but achieves greater accuracy at the faint end, even with only ten clusters, 
thanks to the precision afforded by the highly complete spectroscopic information.
In the future we hope to extend our sample of clusters and to obtain images
of the more nearby objects in the sample of De Propris et al. (2003), using
wide arrays such as NEWFIRM or Omega2000, to firm up the shape of the faint end 
of the LF.

The most interesting conclusion we derive from these data is the essential
similarity of field and cluster LFs in the $K$-band. This was already noted
by De Propris et al. (2003) and Christlein \& Zabludoff (2003), who showed
that $M^*$ and $\alpha$ for cluster and field galaxies differed only at
the $< 10\%$ level. Here, our comparison is for stellar mass functions, as
probed by the $K$-band light. The slope $\alpha$ is the same as in Cole et
al. (2001), while $M^*$ is about 0.2m brighter, but well within errors.

This suggests that the mass function of galaxies, the integral of the merger
and star formation history of all objects down to the present epoch, does
not depend on environment. With more data we will be able to probe the
dependency (or lack thereof) on cluster properties (X-ray luminosity, velocity
dispersion, Bautz-Morgan type), galaxy 'spectrophotometric' class (based on spectral 
features), morphology (based on our images) and radial distance from the cluster 
centre.

The observed lack of environmental dependency is somewhat surprising: one would
expect that the merger histories, and the star formation histories, of galaxies
in clusters and fields would be very different. Therefore the stellar mass function
should show the results of several Gyrs of differential evolution. The observed
lack of environmental dependence, not only in these $K$-band luminosity functions
but also in previous work based on $B$ and $R$ band data (De Propris et al. 2003,
Christlein \& Zabludoff 2003) suggests that either early types dominate the shape
of the LF for galaxies more massive than about 20\% of $L^*$, or that the stellar
populations of such galaxies are predominantly old.

However, it is possible there is a conspiracy between mergers and star formation in producing
nearly similar mass distributions in very different environments. Conversely, the
evidence that most massive galaxies formed early (`downsizing' - Perez-Gonzalez et
al. 2008), argues that the two mass function would converge to similar values at
the present epochs, and that only dwarfs would show significant cluster to field
variation. Observations at higher redshift are needed to explore this issue further.
Nevertheless, a solid understanding of the local situation, as we are trying to 
achieve in this project, is paramount to our interpretation and assessment of high
redshift data.


\begin{thebibliography}{}
  \bibitem{} Balogh, M., Eke, V., Miller, C. J., et al.: 2004, MNRAS, 348, 1335
  \bibitem{} Bell, E. F., de Jong, R.: 2001, ApJ, 550, 212
  \bibitem{} Bertin, E., Arnouts, S.: 1996, A\&AS, 117, 393
  \bibitem{} Christlein, D., Zabludoff, A. I.: 2003, ApJ, 591, 764
  \bibitem{} Cole, S., Norberg, P., Baugh, C. M. et al.: 2001, MNRAS, 326, 255
  \bibitem{} Colless, M. M.: 1989, MNRAS, 237, 799
  \bibitem{} Colless, M., Dalton, G., Maddox, S. et al.: 2001, MNRAS, 328, 1039
  \bibitem{} De Propris, R., Eisenhardt, P. R., Stanford, S. A., Dickinson, M.: 1998, ApJ, 503, L45 
  \bibitem{} De Propris, R., Colless, M. M., Driver, S. P., et al.: 2003, MNRAS, 342, 725
  \bibitem{} De Propris, R., Colless, M. M., Peacock, J. A., et al.: 2004, MNRAS, 351, 125
  \bibitem{} Dressler, A.: 1980, ApJ, 236, 351
  \bibitem{} Gavazzi, G., Pierini, D., Boselli, A.: 1996, A\&A, 312, 397
  \bibitem{} Gomez, P. L., Nichol, R. C., Miller, C. J., et al: 2003, ApJ, 584, 210
  \bibitem{} Hambly, N. C., MacGillivray, H. T., Read, M. A., et al.: 2001, MNRAS, 326, 1279
  \bibitem{} Jarrett, T.-H., Chester, T., Cutri, R., et al.: 2000, AJ, 119, 2498
  \bibitem{} Lewis, I., Balogh, M., De Propris, R., et al.: 2002, MNRAS, 334, 673
  \bibitem{} Lin, Y.-T., Mohr, J. J., Stanford, S. A.: 2004, ApJ, 610, 745
  \bibitem{} P{\`e}rez-Gonz{\'a}lez, P. G., Rieke, G. H., Villar, V. et al.: 2008, ApJ, 675, 234
  \bibitem{} Popesso, P., Biviano, A., B{\"o}hringer, H., Romaniello, M.: 2006, A\&A, 445, 29
\end{thebibliography}
\end{document}